\documentclass[11pt]{article}

\usepackage{fullpage}
\usepackage{latexsym}

\newtheorem{theorem}{Theorem}
\newtheorem{lemma}{Lemma}
\newtheorem{corollary}{Corollary}
\newcommand{\st}[1]{|#1\rangle} % state | >
\newcommand{\lb}[1]{\parallel#1\parallel} % L2-norm || ||
\newcommand{\ol}[1]{\overline{#1}}
\newcommand{\olo}{\overline{0}}

\def\01{\{0,1\}} % binary values

\newenvironment{proof}
{\noindent {\bf Proof }}
{{\hfill $\Box$}\\
 \smallskip}

\begin{document}

\title{Lower Bounds for Quantum Search and Derandomization}
\author{Harry Buhrman\thanks{CWI, P.O.~Box 94709, Amsterdam, 
The Netherlands. E-mail: {\tt buhrman@cwi.nl}.}
\and
Ronald de Wolf\thanks{CWI and University of Amsterdam. 
E-mail: {\tt rdewolf@cwi.nl}.}}
\maketitle

\begin{abstract}
We prove lower bounds on the error probability of a quantum algorithm
for searching through an unordered list of $N$ items, 
as a function of the number $T$ of queries it makes.
In particular, if $T\in O(\sqrt{N})$ then the error is lower bounded by a 
constant. If we want error $\leq 1/2^N$ then we need $T\in\Omega(N)$ queries.
We apply this to show that a quantum computer cannot do much better than
a classical computer when amplifying the success probability of an RP-machine.
A classical computer can achieve error $\leq 1/2^k$ using $k$ applications 
of the RP-machine, a quantum computer still needs at least $ck$ applications 
for this (when treating the machine as a black-box), where $c>0$ is a constant
independent of $k$.
Furthermore, we prove a lower bound of $\Omega(\sqrt{\log N}/\log\log N)$ 
queries for quantum bounded-error search of an {\em ordered} list of $N$ items.
\end{abstract}

\section{Introduction}

Suppose we have an unsorted list of $N$ items and we want to find an item 
with some specific property. 
For instance we want to find an item with a specific value at one of its fields.
In the worst case, a classical deterministic or randomized computer will have 
to look at $\Theta(N)$ items to have a high probability of finding such an 
item if there is one.
On the other hand, Grover's quantum search algorithm can perform look-ups
or queries in superposition, and finds the desired item with high probability 
using only $O(\sqrt{N})$ queries. 
The following is known about the error probability $\varepsilon$ in quantum search:
\begin{itemize}
\item $\varepsilon$ can be made an arbitrarily small constant using $O(\sqrt{N})$ 
queries~\cite{grover:search} but not using $o(\sqrt{N})$ queries
\cite{bbbv:str&weak,bbht:bounds,zalka:grover,bbcmw:polynomials,grover:lower}.
\item $\varepsilon$ can be made $\leq 1/2^{N^\alpha}$ using $O(N^{0.5+\alpha})$ 
queries~\cite[Theorem~1.16]{BuhrmanCleveWigderson98}.
\item If we want no error at all ($\varepsilon=0$), then we need $N$ 
queries~\cite[Corollary~6.2]{bbcmw:polynomials}.
\end{itemize}
Many applications of quantum computing will need to apply quantum 
search several times as a subroutine. 
We should avoid that the errors of each application add
up to an overall error that is too big. Accordingly, we should make
the error probability of each application as small as possible, 
if necessary by spending slightly more than $O(\sqrt{N})$ queries.

We give a detailed analysis of the trade-off between the error 
probability of a quantum search algorithm and the number of queries it uses.
We obtain the following lower bound on $\varepsilon$ in terms of the number 
$T$ of queries that the algorithm uses:
$$
\varepsilon\in\Omega\left(e^{-4bT^2/N-8T/\sqrt{N}}\right),
$$
where $b$ is some fixed constant and we assume $T<N$.
Our proof first translates a quantum search algorithm with $T$ queries to
a multivariate polynomial of degree $d\leq 2T$ that has certain properties, 
and then uses techniques from~\cite{paturi:degree} 
and~\cite{coppersmith&rivlin:poly} to prove a lower bound on $\varepsilon$
in terms of $d$. This implies a lower bound in terms of $T$.%
\footnote{Nayak and Wu~\cite{nayak&wu:mean} also use polynomial-techniques 
from \cite{bbcmw:polynomials} and~\cite{paturi:degree}, in order to prove 
lower bounds for quantum computing the median and mean of a function.}
In particular, this bound implies that $\varepsilon$ cannot be made $o(1)$
using only $O(\sqrt{N})$ queries. Also, $\varepsilon$ can only be
made $\leq 1/2^N$ using $\Omega(N)$ queries.

In Section~\ref{secrp} we apply this bound to the derandomization
of classical RP-machines (RP is the class of languages that can
be recognized in polynomial time with one-sided error at most $1/2$).
A classical computer can achieve error $\leq 1/2^k$ by running the RP-machine
$k$ times and answering `yes' iff at least one of those $k$ runs answered `yes'.
Since this is basically a search among $k$ items, we would expect
a quantum computer to be able to achieve error $\leq 1/2^k$
using roughly $\sqrt{k}$ applications of the RP-machine.
Somewhat surprisingly, we show that a quantum computer 
can{\em not} do much better than the classical computer:
it would also need at least $ck$ applications of the machine to obtain
error $\leq 1/2^k$ (when treating the machine as a black-box).
Here $c>0$ does not depend on $k$.
We interpret this as follows: general results on amplitude amplification
\cite{bht:counting,grover:framework,mosca:eigen} show that a quantum 
computer can achieve a {\em square-root} speed-up when amplifying a very
small success probability to a constant one, but our result shows that
it can achieve at most a {\em linear} speed-up when amplifying 
a constant success probability to a probability very close to 1.

Finally, in Section~\ref{secorderedsearch} we look at the problem of
searching an {\em ordered} list of $N$ items (ordered according to some
key field of the items). Since many databases in practice are ordered
rather than unordered, we feel this problem merits as much attention as
the unordered search has received so far in the quantum computing literature.
Classically, we can search such an ordered list with only $\log N$
queries using binary search. It is unknown whether a quantum computer
can improve on this. However, we show that it cannot improve much more than 
a square-root: we prove a lower bound of $\Omega(\sqrt{\log N}/\log\log N)$ 
queries for bounded-error quantum search in this setting, using
a novel kind of quantum reduction from the PARITY-problem.

To summarize:
\begin{itemize}
\item We prove a general lower bound on the error in quantum search
of an unordered list.
\item We apply this to show that a quantum computer can
achieve at most a linear speed-up when amplifying an already-big
success probability.
\item We prove a lower bound of roughly $\sqrt{\log N}$ for
quantum search of an ordered list of $N$ items.
\end{itemize}

\section{Preliminaries}

In this section we define the setting of quantum gate networks (which are
equivalent to quantum Turing machines~\cite{yao:qcircuit}) and queries.

A {\em qubit} is a superposition $\alpha_0\st{0}+\alpha_1\st{1}$ of both values
of a classical bit.
Similarly, a register of $m$ qubits is a superposition $\st{\phi}$
of all $2^m$ classical bitstrings of $m$ bits, written
$$
\st{\phi}=\sum_{k\in\{0,1\}^m}\alpha_k\st{k}.
$$
Here $\alpha_k$ is a complex number, called the {\em amplitude} of state $\st{k}$.
If we observe $\st{\phi}$ we will see one and only one $\st{k}$.
The probability of seeing one specific $\st{k}$ is given by $|\alpha_k|^2$.
Hence we must have $\sum_{k\in\{0,1\}^m}|\alpha_k|^2=1$.
After observing $\st{\phi}$ and seeing $\st{k}$, 
the superposition $\st{\phi}$ has collapsed to $\st{k}$.

If we do not observe a state, quantum mechanics tells us that it will evolve
unitarily. This means that the vector of amplitudes is transformed according
to a linear operator that preserves norm (so the sum of the amplitudes squared
remains 1). A unitary operator $U$ always has an inverse $U^{-1}$, 
which equals its conjugate transpose $U^*$.
A quantum gate network working on $m$ qubits 
is like a classical circuit working on $m$ classical bits, except
that instead of AND-, OR-, and NOT-gates we have quantum gates which operate
unitarily on one or more qubits. 
A quantum gate network transforms an initial state into a final state much in 
the way a classical circuit transforms its input into one or more output bits.
It is known that operations on one or two qubits at a time are sufficient to
build any unitary transformation~\cite{barencoea:gates}.
The most common measure of complexity of a quantum gate network is
the number of elementary quantum gates it contains, but in this paper
we will disregard this and only count the number of queries.

Making queries to a list $X=(x_0,\ldots,x_{N-1})$ of $N$ bits is incorporated 
in the model as follows. Classically, making a query to $X$ means 
inputting some $j\in\{0,\ldots,N-1\}$ into a black-box, 
and receiving the value $x_j$ as output.
A query gate $O$ (for ``oracle'') performs the corresponding mapping, 
which is our only way to access the bits $x_j$:
$$
\st{j,0,\olo}\rightarrow\st{j,x_j,\olo},
$$
where $\olo$ is a string of zeroes.
Because $O$ must be reversible, it also maps
$$
\st{j,1,\olo}\rightarrow\st{j,\ol{x_j},\olo}.
$$
We will look at quantum networks that contain both elementary gates and 
query gates, but only count the latter.
The advantage of a quantum computer over a classical computer is its
ability to  make queries in superposition: applying $O$ once to the state
$\frac{1}{\sqrt{N}}\sum_j\st{j,0,\ol{0}}$ results in 
$\frac{1}{\sqrt{N}}\sum_j\st{j,x_j,\ol{0}}$, 
which in some sense ``contains'' all the bits $x_j$.

In terms of linear algebra, a quantum gate network $A$ with $T$ queries can 
be viewed as follows: first $A$ applies some unitary operation $U_0$ to the
initial state, then it applies $O$, then it applies another $U_1$, 
another $O$, and so on up till $U_T$.
Thus $A$ corresponds to a big unitary transformation
$$
A=U_TOU_{T-1}O\ldots OU_1OU_0.
$$
The behavior of $O$ depends on $X$, but the $U_i$ are fixed unitary transformations
independent of $X$. 
We fix the initial state to $\st{\ol{0}}$, independent of $X$. 
The final state is then a superposition $A\st{\ol{0}}$ 
which depends on $X$ only via the $T$ query gates.

One specific bit of the final state (the rightmost one, say) is considered
the output bit.  The output of the network is defined as the value we 
obtain if we observe this bit. Note that the output is a random variable.
The acceptance probability of a quantum network on a specific black-box $X$
is defined to be the probability that the output is 1.
The key lemma of~\cite{bbcmw:polynomials} gives the following relation
between a $T$-query network and a polynomial that expresses its acceptance
probability as a function of $X$ (such a relation is also implicit in some 
of the proofs of \cite{fortnow&rogers:limitations,ffkl:toolkit}):

\begin{lemma}
The acceptance probability of a quantum network that makes $T$ queries to 
a black-box $X$, can be written as a real-valued multilinear $N$-variate 
polynomial $P(X)$ of degree at most $2T$.
\end{lemma}

Note that if we want to compute a Boolean function, then the acceptance probability
$P(X)$ should be close to 1 if $f(X)=1$, and $P(X)$ should be close to 0 if $f(X)=0$.
Since the degree of $P$ is $\leq 2T$, a lower bound on the degree of a polynomial 
with such properties implies a lower bound on $T$. 
See~\cite{bbcmw:polynomials} for some of the lower
bounds on quantum query complexity that can be obtained in this way.

An $N$-variate polynomial $P$ of degree $d$ can be reduced to a single-variate 
one in the following way (due to~\cite{minsky&papert:perceptrons}).
Let $P^{sym}$ be the polynomial that averages $P$ over all permutations of its input:
$$
P^{sym}(X)=\frac{\sum_{\pi\in S_N}P(\pi(X))}{N!}.
$$
$P^{sym}$ is an $N$-variate polynomial of degree at most $d$.
It can be shown that there is a single-variate polynomial $Q$ of degree at most $d$,
such that $P^{sym}(X)=Q(|X|)$ for all $X\in\{0,1\}^N$.
Here $|X|$ denotes the Hamming weight (number of 1s) of $X$.

\section{Lower Bound on the Error in Quantum Search}

In this section we establish a general lower bound on the error probability
in quantum search. Consider an unordered 
list of $N$ items. We will abstract from the
specific contents of the items, treating the list like a kind of black-box.
A query at place $j$ of the list just returns one bit $x_j$, indicating
whether the $j$th item on the list has the property we are looking for.
A query gate performs the following mapping, which is our only
access to the bits $x_j$:
$$
\st{j,b,\olo}\rightarrow\st{j,b\oplus x_j,\olo},
$$
where $b$ is a bit and $\olo$ is a string of zeroes.
The aim is to find a $j$ such that $x_j=1$ (if there is one), 
using as few queries as possible.

Rather than proving a lower bound on search directly, we will prove a
lower bound on computing the OR-function (i.e.\ determining whether $X$
contains at least one 1). This clearly reduces to search.
The main idea of our proof is the following. 
By the lemma of the previous section, the acceptance probability of 
a quantum computer with $T$ queries that computes the OR with error probability
$\leq\varepsilon$ can be written as a multivariate polynomial
of degree $\leq 2T$ of the $N$ bits in the list.
This polynomial can be reduced to a single-variate polynomial $s$ 
of degree $d\leq 2T$ with the following properties: 
\begin{quote}
$s(0)=0$ %
\footnote{Since we can always test whether we actually found a solution
at the expense of one more query, we can assume the algorithm always
gives the right answer `no' if the list contains only 0s.
Hence $s(0)=0$. However, our results remain unaffected if we allow
a small error here also (i.e.\ $0\leq s(0)\leq\varepsilon$).}\\
$1-\varepsilon\leq s(x)\leq 1$ for all integers $x\in[1,N]$
\end{quote}
We will prove a lower bound on $\varepsilon$ in terms of $d$, 
which implies a lower bound in terms of $T$.
Because we can achieve $\varepsilon=0$ iff $T=N$
\cite[Proposition~6.1]{bbcmw:polynomials}, we assume $T<N$
and hence $\varepsilon>0$.

Define $p(x)=1-s(N-x)$. Then $p$ has degree $d$ and
\begin{quote}
$0\leq p(x)\leq\varepsilon$ for all integers $x\in[0,N-1]$\\
$p(N)=1$
\end{quote}
Thus $p$ is ``small'' at integer points in $[0,N-1]$ and ``big'' at $N$.
Coppersmith and Rivlin~\cite[p.~980]{coppersmith&rivlin:poly} 
prove the following theorem, which allows us to show that $p$ is also
``small'' at {\em non}-integer points in $[0,N-1]$.

\begin{theorem}[Coppersmith \&\ Rivlin]
There exist positive constants $a$ and $b$ with the following property.
For every polynomial $p$ of degree $d$ such that 
$$
|p(x)|\leq 1 \mbox{ for all integers } x\in[0,n]
$$ 
and any $\delta>0$ such that $n\geq\delta d^2$, we have
$$
|p(x)|< a e^{b/\delta} \mbox{ for all real } x\in[0,n].
$$
\end{theorem}

Let $\delta=(N-1)/d^2$.
Applying Coppersmith and Rivlin's theorem to $p/\varepsilon$ 
(which is bounded by 1 at integer points) we obtain:
$$
|p(x)|<\varepsilon a e^{b/\delta} \mbox{ for all real } x\in[0,N-1].
$$
Now we rescale $p$ to $q(x)=p((x+1)(N-1)/2)$
(i.e. the domain $[0,N-1]$ is transformed to $[-1,1]$), which has the
following properties:
\begin{quote}
$|q(x)|<\varepsilon a e^{b/\delta} \mbox{ for all real } x\in[-1,1]$\\
For $\mu=2/(N-1)$ we have $q(1+\mu)=p(N)=1$
\end{quote}
Thus $q$ is ``small'' on all $x\in[-1,1]$ and ``big'' just outside this interval
($q(1+\mu)=1$).

Let $T_d$ denote the degree-$d$ Chebyshev polynomial~\cite{rivlin:chebyshev}:
$$
T_d(x)=\frac{1}{2}
\left(\left(x+\sqrt{x^2-1}\right)^d+\left(x-\sqrt{x^2-1}\right)^d\right).
$$
The following is known:
\begin{itemize}
\item If $q$ is a polynomial of degree $d$ such that
$|q(x)|\leq c$ for all $x\in[-1,1]$ then $|q(x)|\leq c|T_d(x)|$ for all $|x|\geq 1$
\cite[Fact~2]{paturi:degree}\cite[p.108]{rivlin:chebyshev}
\item $T_d(1+\mu)\leq e^{2d\sqrt{2\mu+\mu^2}}$ for all $\mu\geq 0$
\cite[p.471, before Fact~2]{paturi:degree}\footnote{For $x=1+\mu$:
$T_d(x)\leq (x+\sqrt{x^2-1})^d=(1+\mu+\sqrt{2\mu+\mu^2})^d\leq
(1+2\sqrt{2\mu+\mu^2})^d \leq e^{2d\sqrt{2\mu+\mu^2}}$ 
(Paturi, personal communication).}
\end{itemize}
Linking all this we obtain
$$
1=q(1+\mu) 
\leq \varepsilon a e^{b/\delta} |T_d(1+\mu)|
\leq \varepsilon a e^{b/\delta+2d\sqrt{2\mu+\mu^2}}.
$$
This shows that if $q$ is ``big'' just outside the interval $[-1,1]$, 
then it cannot have been very small inside this interval, so $\varepsilon$
cannot have been very small.
Substituting $\delta=(N-1)/d^2$ and $\mu=2/(N-1)$ 
we obtain the following lower bound on $\varepsilon$:
$$
\varepsilon\geq\frac{1}{a}e^{-bd^2/(N-1)-4d/\sqrt{N/(N-1)^2}}.
$$
Since $d\leq 2T$, where $T$ is the number of queries of 
the quantum search algorithm, we have (simplifying a bit):

\begin{theorem}
If $T<N$ then
$\displaystyle\varepsilon\in\Omega\left(e^{-4bT^2/N-8T/\sqrt{N}}\right)$.
\end{theorem}

We note some special cases of this general theorem:

\begin{corollary}
No quantum network for bounded-error search of an unordered list
that uses $O(\sqrt{N})$ queries can have error probability $o(1)$.
\end{corollary}

For instance, an error $\leq 1/N$ cannot be achieved 
using only $O(\sqrt{N})$ queries.\footnote{Which is too bad, because such 
a small error would reduce the quantum complexity of $\Sigma_2$ 
(the second level of the polynomial hierarchy) from $O(\sqrt{2^n}n)$ 
to $O(\sqrt{2^n})$~\cite{BuhrmanCleveWigderson98}.}

\begin{corollary}
Every quantum network for bounded-error search of an unordered list that
uses $\leq N^{0.5+\alpha}$ queries ($\alpha\geq 0$) must have error probability 
$\Omega\left(1/2^{cN^{2\alpha}}\right)$ (where $c>0$ is some fixed constant).
\end{corollary}

In particular, this shows that we cannot obtain error probability $\leq 1/2^N$
unless we have $\alpha=0.5$ and thus use $\Omega(N)$ queries.
\cite[Theorem~1.16]{BuhrmanCleveWigderson98} proves the upper bound that
the error probability can be made as small as $1/2^{N^{\alpha}}$ using
$O(N^{0.5+\alpha})$ queries, so there is still a gap between upper and lower bound.

Finally, a lower bound on $T$ in terms of $\varepsilon$ and $N$:

\begin{corollary}
If $T(N)/\sqrt{N}\rightarrow\infty$ but $T<N$, then 
$\displaystyle T\in\Omega\left(\sqrt{N\log(1/\varepsilon)}\right)$.
\end{corollary}

\section{The Influence of the Number of Solutions}

Suppose we have a quantum search algorithm that uses $T$ queries and
works well (i.e.\ has error $\leq\varepsilon$) whenever the number of 1s 
in the list of $N$ items is either 0 or at least $t$.  
(Here $t$ is some fixed number $<N$.)
Such an algorithm induces a polynomial of degree $d\leq 2T$ 
with the following properties:
\begin{quote}
$s(0)=0$\\
$1-\varepsilon\leq s(x)\leq 1$ for all integers $x\in[t,N]$
\end{quote}
Define $p(x)=1-s(N-x)$, which has degree $d$ and
\begin{quote}
$0\leq p(x)\leq\varepsilon$ for all integers $x\in[0,N-t]$\\
$p(N)=1$
\end{quote}
Now we define $q(x)=p((x+1)(N-t)/2)$, $\delta=(N-t)/d^2$ and $\mu=2t/(N-t)$, 
and derive completely analogous to the previous section:
\begin{eqnarray*}
1 & = & q(1+\mu) 
\leq \varepsilon a e^{b/\delta} |T_d(1+\mu)|
\leq \varepsilon a e^{b/\delta+2d\sqrt{2\mu+\mu^2}}\\
 & = &\varepsilon a e^{bd^2/(N-t)+2d\sqrt{4t/(N-t)+4t^2/(N-t)^2}}\\
 & = &\varepsilon a e^{bd^2/(N-t)+4d\sqrt{tN/(N-t)^2}}.
\end{eqnarray*}
Hence for quantum search in this situation we have the bound:
$$
\varepsilon\in\Omega\left(e^{-bd^2/(N-t)-4d\sqrt{tN/(N-t)^2}}\right)
\in\Omega\left(e^{-4bT^2/(N-t)-8T\sqrt{tN/(N-t)^2}}\right).
$$
\cite{bbht:bounds} proves that an expected number of $O(\sqrt{N/t})$
queries is sufficient to search with high probability.
If we put $T=c\sqrt{N/t}$ then the lower bound on the error probability
becomes roughly $\Omega(e^{-c'c})$ (for some constant $c'>0$), which can
indeed be made arbitrarily small by increasing $c$.
On the other hand, if $T\in o(\sqrt{N/t})$ then the lower bound on the error
goes to the constant $1/a$ for $N\rightarrow\infty$ and $t=o(N)$.
Now if we were able to achieve some error $<1/2$ using $o(\sqrt{N/t})$
queries, we could also make the error $<1/a$ by repeating a constant
number of times, which would still take only $o(\sqrt{N/t})$ queries.
This shows that we can{\em not} achieve error $<1/2$ using $o(\sqrt{N/t})$ 
queries.
Thus the $O(\sqrt{N/t})$ upper bound is tight up to a constant factor 
(as already shown in a different way in~\cite{bbht:bounds}).

\section{Application to Derandomization of RP}\label{secrp}

Let $A$ be some RP-algorithm for a language $L$ with running time $\leq p(n)$.
$A$ always gives the right answer `no' for every input $x\not\in L$, and gives 
the right answer `yes' with probability at least $1/2$ for every $x\in L$.
We want to lower the error probability using as few calls to $A$ as possible.
For a fixed input $x$ of length $n$ we can consider $A$ as a black-box of 
$N\leq 2^{p(n)}$ items. Each item corresponds to the value $A$ outputs when given 
a specific random string ($A$ can use at most $p(n)$ random bits and hence
at most $2^{p(n)}$ distinct random strings).
By definition of RP, this black-box satisfies the promise that either 
it contains 0 1s (if $x\not\in L$) or at least $N/2$ 1s (if $x\in L$).

A classical computer can improve the error probability to at most $1/2^k$ by
making $k$ black-box queries (i.e.\ $k$ applications of the algorithm
on $k$ different random strings) and answering `yes' iff at least one
those $k$ queries answered `yes'. 
How much better can a quantum computer do, if we only allow it 
to call $A$ as a black-box?
Note that the classical method basically searches through a list of $k$ items,
looking for a 1. 
Accordingly, the following quantum algorithm suggests itself:
select $k$ random strings and search whether one of these gives a `yes'
in $O(\sqrt{k})$ applications of the algorithm.
Thus we would expect a quantum computer to be able to achieve the same 
error probability $\leq 1/2^k$ using roughly $\sqrt{k}$ applications of the
algorithm instead of $k$.

However, note that the situation here
corresponds exactly to the previous section with $t=N/2$.
Thus if the quantum computer makes $T$ queries and has error probability
$\varepsilon$ on the worst-case black-box, then
$$
\varepsilon\in\Omega\left(e^{-8bT^2/N-8T\sqrt{2}}\right).
$$
If we want $\varepsilon\leq 1/2^k$ (for some fixed $k$ and all $N$), 
it follows that $T\geq ck$, for some $c>0$ that does not depend on $k$.%
\footnote{For sufficiently large $k$,
$c$ will be roughly $1/8\sqrt{2}\log e\approx 0.06$.}
Thus the quantum algorithm can{\em not} achieve the square-root speed-up 
that we expected; it can achieve at most a linear speed-up.

Why does the above-mentioned $\sqrt{k}$-method not work?
The reason is that the quantum searching algorithm itself has some
error probability, in addition to the probability $\leq 1/2^k$ that
the chosen sample of $k$ items does not contain a 1 when the larger
list of $N$ items {\em does} contain a 1. The error introduced by quantum
search can only be made sufficiently small at the cost of increasing
$k$ and/or the number of queries spent.

In sum: on a classical computer we can amplify an RP-algorithm to error 
probability $\varepsilon\leq 1/2^k$ using $k$ applications of the algorithm,
on a quantum computer we cannot do much better: we still need at least
$ck$ applications to achieve error $\varepsilon\leq 1/2^k$,
provided we use the RP-machine only as a black-box.

\section{Lower Bound on Search in an Ordered List}\label{secorderedsearch} 

Grover's algorithm can find a specific item in an unordered list of $N$ 
items with high probability, using only $O(\sqrt{N})$ queries (a.k.a.\ database 
look-ups), whereas a classical algorithm needs $\Theta(N)$ queries for this.
There exist several lower-bound proofs that show that the $O(\sqrt{N})$ is optimal
\cite{bbbv:str&weak,bbht:bounds,zalka:grover,bbcmw:polynomials,grover:lower}.

What about search in a list of $N$ items which is {\em ordered} according to
some key-value of each item?
A classical deterministic algorithm can search such a list using $\log N$
queries by means of binary search (each query can effectively halve the 
relevant part of the list: looking at the key of the middle item of the list 
tells you whether the item you are searching for is in the first or the
second half of the list). 
How much better can we do on a quantum computer?
Can we again get a square-root speed-up?
Here we show that the speed-up cannot be much better than a square-root:
we prove a lower bound of $\Omega(\sqrt{\log N}/\log\log N)$ queries for 
bounded-error quantum search of an ordered list.
In contrast, we have no upper bound better than the classical $\log N$.

We will formalize a query on an ordered list as follows, 
abstracting from the specific contents of the key field.
The list is viewed as a list of $N$ bits, $x_0,\ldots,x_{N-1}$,
and there is an unknown number $i$ such that $x_j=1$ iff $j\leq i$.
Here $x_j$ being 1 can be interpreted as saying that the $j$th item
on the list has a key-value smaller or equal to the value we are looking for.
The goal is to find the number $i$, which is the point in the list where
the looked-for item resides, using as few queries as possible.
In quantum network terms, a query corresponds to a gate $C$ that maps
$$
\st{j,b,\olo}\rightarrow\st{j,b\oplus x_j,\olo}.
$$
The following theorem proves a lower bound of roughly $\sqrt{\log N}$ queries
for quantum searching an ordered list with bounded error probability.
To improve readability, we have deferred some of the more technical details 
to the appendix.
Basically these show that we can approximately simulate the gate $C$ 
using roughly $\sqrt{\log N}$ queries to a black-box of $\log N$ bits 
that represents the number $i$.

\begin{theorem}
A quantum network for bounded-error search of an ordered list of $N$ items must use 
at least $\Omega(\sqrt{\log N}/\log\log N)$ queries.
\end{theorem}

\begin{proof}
Suppose we have a network $S$ for bounded-error ordered search that uses $T$ queries
to find the number $i$ hidden in an ordered black-box $X$ with high probability.
Since $\log N$ queries are sufficient for this (classical binary search),
we can assume $T\leq\log N$.
We will show how we can get from $S$ to a network $\widetilde{S}$ that 
determines the whole contents of an arbitrary black-box $Y$ of $\log N$ bits 
with high probability, using only $T\cdot O(\sqrt{\log N}\log\log N)$ queries 
to $Y$. This would allow us to compute the PARITY-function of $Y$ (i.e.\
whether or not $Y$ contains odd many 1s).
Since we have a $(\log N)/2$ lower bound for 
the latter~\cite[Proposition~6.4]{bbcmw:polynomials}, we have
$$
T\cdot O(\sqrt{\log N}\log\log N)\geq\frac{\log N}{2},
$$
from which the theorem follows.

So let $Y$ be an arbitrary black-box of $\log N$ bits.
This represents a number $i\in\{0,\ldots,N-1\}$.
Let $X=(x_0,\ldots,x_{N-1})$ be the ordered black-box corresponding to $i$,
so $x_j=1$ iff $j\leq i$.
The network $S$, when allowed to make queries to $X$, 
outputs the number $i$ with high probability.
A query-gate $C$ for $X$ maps 
$$
\st{j,b,\olo}\rightarrow\st{j,b\oplus x_j,\olo}.
$$
Since $x_j=1$ iff $j\leq i$, Lemmas~\ref{lemgeq} and~\ref{lemcleanup} of 
the appendix imply that there is a quantum network $\widetilde{C}$ that
uses $O(\sqrt{\log N}\log\log N)$ queries to $Y$ and maps
$$
\st{j,b,\olo}\rightarrow\st{j,b\oplus x_j,\olo}+\st{j}\st{W_{jb}},
$$
where $\lb{\st{W_{jb}}}\leq\eta/\log N$ for all $j,b$, 
for some small fixed $\eta$ of our choice.

Let $\widetilde{S}$ be obtained from $S$ by replacing all $T$ $C$-gates 
by $\widetilde{C}$-networks. Note that
$\widetilde{S}$ contains $T\cdot O(\sqrt{\log N}\log\log N)$ queries to $Y$.
Consider the way $\widetilde{S}$ acts on initial state $\st{\olo}$, compared to $S$. 
Each replacement of $C$ by $\widetilde{C}$ introduces an error, but each of these 
errors is at most $\sqrt{2}\eta/\log N$ in Euclidean norm by Lemma~\ref{lemeucldist}.
By unitarity these $T$ errors add linearly, so the final states will be close together:
$$
\lb{S\st{\olo}-\widetilde{S}\st{\olo}}\leq T\sqrt{2}\eta/\log N\leq\sqrt{2}\eta.
$$
Since observing the final state $S\st{\olo}$ yields the number $i$ with high probability,
observing $\widetilde{S}\st{\olo}$ will also yield $i$ with high probability.
Thus the network $\widetilde{S}$ allows us to learn $i$, and hence the whole black-box $Y$.
\end{proof}
\section*{Acknowledgements}
We would like to thank David Deutsch, Wim van Dam and Mike Mosca for
discussions which emphasized the importance of making the error in
quantum search as small as possible.

%\bibliography{qc}

\begin{thebibliography}{CDNT97}

\bibitem[BBBV97]{bbbv:str&weak}
C.~H. Bennett, E.~Bernstein, G.~Brassard, and U.~Vazirani.
\newblock Strengths and weaknesses of quantum computing.
\newblock {\em SIAM Journal on Computing}, 26(5):1510--1523, 1997.
\newblock quant-ph/9701001.

\bibitem[BBC{\etalchar{+}}95]{barencoea:gates}
A.~Barenco, C.H. Bennett, R.~Cleve, D.P. DiVincenzo, N.~Margolus, P.~Shor,
  T.~Sleator, J.~Smolin, and H.~Weinfurter.
\newblock Elementary gates for quantum computation.
\newblock {\em Physical Review A}, 52:3457--3467, 1995.

\bibitem[BBC{\etalchar{+}}98]{bbcmw:polynomials}
R.~Beals, H.~Buhrman, R.~Cleve, M.~Mosca, and R.~de Wolf.
\newblock Quantum lower bounds by polynomials.
\newblock In {\em Proceedings of 39th FOCS}, pages 352--361, 1998.
\newblock also quant-ph/9802049.

\bibitem[BBHT98]{bbht:bounds}
M.~Boyer, G.~Brassard, P.~H{\o}yer, and A.~Tapp.
\newblock Tight bounds on quantum searching.
\newblock {\em Fortschritte der Physik}, 46(4--5):493--505, 1998.
\newblock Earlier version in Physcomp'96; also quant-ph/9605034.

\bibitem[BCW98]{BuhrmanCleveWigderson98}
H.~Buhrman, R.~Cleve, and A.~Wigderson.
\newblock Quantum vs.\ classical communication and computation (preliminary
  version).
\newblock In {\em Proceedings of 30th STOC}, pages 63--68, 1998.
\newblock quant-ph/9802040.

\bibitem[BHT98]{bht:counting}
G.~Brassard, P.~H{\o}yer, and A.~Tapp.
\newblock Quantum counting.
\newblock In {\em Proceedings of 25th ICALP}, volume 1443 of {\em Lecture Notes
  in Computer Science}, pages 820--831. Springer, 1998.
\newblock quant-ph/9805082.

\bibitem[CDNT97]{cdnt:ip}
R.~Cleve, {W. van} Dam, M.~Nielsen, and A.~Tapp.
\newblock Quantum entanglement and the communication complexity of the inner
  product function.
\newblock quant-ph/9708019, 10 Aug 1997.

\bibitem[CR92]{coppersmith&rivlin:poly}
D.~Coppersmith and T.~J. Rivlin.
\newblock The growth of polynomials bounded at equally spaced points.
\newblock {\em SIAM Journal on Mathematical Analysis}, 23(4):970--983, 1992.

\bibitem[DH96]{durr&hoyer:minimum}
C.~D{\"u}rr and P.~H{\o}yer.
\newblock A quantum algorithm for finding the minimum.
\newblock quant-ph/9607014, 18 Jul 1996.

\bibitem[FFKL93]{ffkl:toolkit}
S.~Fenner, L.~Fortnow, S.~Kurtz, and L.~Li.
\newblock An oracle builder's toolkit.
\newblock In {\em Proceedings of the 8th {IEEE} Structure in Complexity Theory
  Conference}, pages 120--131, 1993.

\bibitem[FR98]{fortnow&rogers:limitations}
L.~Fortnow and J.~Rogers.
\newblock Complexity limitations on quantum computation.
\newblock In {\em Proceedings of the 13th {IEEE} Conference on Computational
  Complexity}, pages 202--209, 1998.

\bibitem[Gro96]{grover:search}
L.~K. Grover.
\newblock A fast quantum mechanical algorithm for database search.
\newblock In {\em Proceedings of 28th STOC}, pages 212--219, 1996.
\newblock quant-ph/9605043.

\bibitem[Gro98a]{grover:lower}
L.~K. Grover.
\newblock How fast can a quantum computer search?
\newblock quant-ph/9809029, 10 Sep 1998.

\bibitem[Gro98b]{grover:framework}
L.~K. Grover.
\newblock A framework for fast quantum mechanical algorithms.
\newblock In {\em Proceedings of 30th STOC}, pages 53--62, 1998.
\newblock quant-ph/9711043.

\bibitem[Mos98]{mosca:eigen}
M.~Mosca.
\newblock Quantum searching, counting and amplitude amplification by
  eigenvector analysis.
\newblock In {\em MFCS'98 workshop on Randomized Algorithms}, 1998.

\bibitem[MP68]{minsky&papert:perceptrons}
M.~Minsky and S.~Papert.
\newblock {\em Perceptrons}.
\newblock MIT Press, Cambridge, MA, 1968.
\newblock Second, expanded edition 1988.

\bibitem[NW98]{nayak&wu:mean}
A.~Nayak and F.~Wu.
\newblock On the quantum black-box complexity of approximating the mean and the
  median.
\newblock quant-ph/9804066, 29 Apr 1998.

\bibitem[Pat92]{paturi:degree}
R.~Paturi.
\newblock On the degree of polynomials that approximate symmetric {B}oolean
  functions (preliminary version).
\newblock In {\em Proceedings of 24th STOC}, pages 468--474, 1992.

\bibitem[Riv90]{rivlin:chebyshev}
T.~J. Rivlin.
\newblock {\em Chebyshev Polynomials: From Approximation Theory to Algebra and
  Number Theory}.
\newblock Wiley-Interscience, second edition, 1990.

\bibitem[Yao93]{yao:qcircuit}
A.~C-C. Yao.
\newblock Quantum circuit complexity.
\newblock In {\em Proceedings of 34th FOCS}, pages 352--360, 1993.

\bibitem[Zal97]{zalka:grover}
C.~Zalka.
\newblock Grover's quantum searching algorithm is optimal.
\newblock quant-ph/9711070, 26 Nov 1997.

\end{thebibliography}

\newcommand{\etalchar}[1]{$^{#1}$}

\appendix

\section{Some Technical Lemmas}

Our lower-bound proof for ordered search uses three technical lemmas.

The first lemma can be obtained from the result of D{\"u}rr and 
H{\o}yer~\cite{durr&hoyer:minimum} that a quantum algorithm can find 
the {\em minimum} element on a list of $N$ items using $O(\sqrt{N})$ queries.
We can use this to find the leftmost bit where two lists differ, which
tells us which of the two numbers represented by the two lists is bigger.

\begin{lemma}\label{lemgeq}
There exists a quantum algorithm $A$ that with bounded error probability
outputs on input $j$ ($0\leq j \leq N-1$) whether $j$ is smaller or 
equal to a number $i$ represented by a black-box of $\log N$ bits,
using $O(\sqrt{\log N})$ queries to the black-box.
\end{lemma}
 
By standard techniques, we can make the error probability $O(1/\log N)$
by repeating the algorithm $O(\log\log N)$ times.

The second lemma shows how to obtain an approximately ``clean'' computation
that uses no measurements (the proof is as in~\cite[Section~3]{cdnt:ip} and
\cite[Theorem~1.14]{BuhrmanCleveWigderson98}).

\begin{lemma}\label{lemcleanup}
Suppose there exists a quantum algorithm $A$ that uses $T$ queries and 
outputs a bit $x_j$ with error probability $\leq\varepsilon$ on initial 
state $\st{j,\olo}$, for every $j$, and does not change the $j$-register.
Then there exists a quantum algorithm $A'$ that uses $2T$ queries
and no measurements, and maps
$$
\st{j,b,\olo}\rightarrow\st{j,b\oplus x_j,\olo}+\st{j}\st{W_{jb}},
$$
where $\lb{\st{W_{jb}}}\leq\sqrt{2\varepsilon}$, for every $j$ and $b\in\{0,1\}$.
\end{lemma}

\begin{proof}
The idea is the familiar ``compute, copy answer, uncompute''-sequence.
By standard techniques, we can assume $A$ itself uses no measurements
and is followed by a single measurement.
Then there exist amplitudes $\alpha_0$ and $\alpha_1$
and unit-length vectors $\st{V_0}$ and $\st{V_1}$ such that
$$
A\st{j,0,\olo}=\alpha_0\st{j,x_j}\st{V_0}+\alpha_1\st{j,\ol{x_j}}\st{V_1},
$$
and $|\alpha_1|^2\leq\varepsilon$.
For ease of notation, we assume this state is preceded by the bit $b$.
Applying the controlled-not operation that maps 
$\st{b,j,x}\rightarrow\st{b\oplus x,j,x}$, we get
$$
\alpha_0\st{b\oplus x_j,j,x_j}\st{V_0}+\alpha_1\st{b\oplus\ol{x_j},j,\ol{x_j}}\st{V_1}=
$$
$$
\st{b\oplus x_j}\left(\alpha_0\st{j,x_j}\st{V_0}+\alpha_1\st{j,\ol{x_j}}\st{V_1}\right)+
\alpha_1\st{b\oplus\ol{x_j},j,\ol{x_j}}\st{V_1}-\alpha_1\st{b\oplus x_j,j,\ol{x_j}}\st{V_1}.
$$
Applying $I\otimes A^{-1}$ gives
$$
\st{b\oplus x_j}\st{j,0,\olo}+(I\otimes A^{-1}) \left(\alpha_1\st{b\oplus\ol{x_j},j,\ol{x_j}}\st{V_1}-
\alpha_1\st{b\oplus x_j,j,\ol{x_j}}\st{V_1}\right).
$$
Applying an operation $B$ which swaps the first bit and $j$, we get
$$
\st{j,b\oplus x_j,0,\olo}+B(I\otimes A^{-1}) \left(\alpha_1\st{b\oplus\ol{x_j},j,\ol{x_j}}\st{V_1}-
\alpha_1\st{b\oplus x_j,j,\ol{x_j}}\st{V_1}\right).
$$
Note that $B(I\otimes A^{-1}) \left(\alpha_1\st{b\oplus\ol{x_j},j,\ol{x_j}}\st{V_1}-
\alpha_1\st{b\oplus x_j,j,\ol{x_j}}\st{V_1}\right)=\st{j}\st{W_{jb}}$ for some $\st{W_{jb}}$,
because $A$ and hence also $A^{-1}$ do not change $j$.
Now
\begin{eqnarray*}
\lb{\st{W_{jb}}} & = & \lb{\st{j}\st{W_{jb}}}\\
                 & = & \lb{B(I\otimes A^{-1})\left(\alpha_1\st{b\oplus\ol{x_j},j,\ol{x_j}}\st{V_1}-
\alpha_1\st{b\oplus x_j,j,\ol{x_j}}\st{V_1}\right)}\\
            & = & \lb{\alpha_1\st{b\oplus\ol{x_j},j,\ol{x_j}}\st{V_1}-\alpha_1\st{b\oplus x_j,j,\ol{x_j}}\st{V_1}}\\
       & = & \sqrt{2|\alpha_1|^2}\leq \sqrt{2\varepsilon}.
\end{eqnarray*}
Thus the quantum algorithm $A'$ which first applies $A$, 
then XORs the answer-bit into $b$, and then applies $A^{-1}$, 
satisfies the lemma.
\end{proof}

The next lemma uses an idea from~\cite{cdnt:ip}.
It shows that if we can simulate a gate $C$ by means of a network 
$\widetilde{C}$ that works well on basis states, then $\widetilde{C}$
also works well on superpositions of basis states.

\begin{lemma}\label{lemeucldist}
Let $C$ and $\widetilde{C}$ be unitary transformations such that
\begin{quote}
$C: \st{j,b,\olo}\rightarrow\st{j,b\oplus x_j,\olo}$\\
$\widetilde{C}: \st{j,b,\olo}\rightarrow\st{j,b\oplus x_j,\olo}+\st{j}\st{W_{jb}}$
\end{quote}
If $\lb{\st{W_{jb}}}\leq\varepsilon$ for every $j\in\{0,\ldots,N-1\}$ and 
$b\in\{0,1\}$, and $\st{\phi}=\sum_{j,b}\alpha_{jb}\st{j,b,\olo}$ has norm 1, then 
\begin{quote}
$\lb{C\st{\phi}-\tilde{C}\st{\phi}}\leq\sqrt{2}\varepsilon$.
\end{quote}
\end{lemma}

\begin{proof}
\begin{eqnarray*}
\lb{C\st{\phi}-\tilde{C}\st{\phi}} & = & \lb{\sum_{j,b} \alpha_{jb}\st{j}\st{W_{jb}}}\\
   & \leq & \lb{\sum_{j} \alpha_{j0}\st{j}\st{W_{j0}}} +
            \lb{\sum_{j} \alpha_{j1}\st{j}\st{W_{j1}}}\\
   & \stackrel{(1)}{=} & \sqrt{\sum_{j}|\alpha_{j0}|^2\lb{\st{j}\st{W_{j0}}}^2} +
            \sqrt{\sum_{j}|\alpha_{j1}|^2\lb{\st{j}\st{W_{j1}}}^2}\\
   & \leq & \varepsilon\sqrt{\sum_{j}|\alpha_{j0}|^2}+
            \varepsilon\sqrt{\sum_{j}|\alpha_{j1}|^2}
            \ \stackrel{(2)}{\leq} \ \sqrt{2}\varepsilon.
\end{eqnarray*}
Here $(1)$ holds because the states $\st{j}\st{W_{jb}}$ in 
$\sum_j\alpha_{jb}\st{j}\st{W_{jb}}$ are all orthogonal,
and $(2)$ holds because $\sqrt{a}+\sqrt{1-a}\leq\sqrt{2}$ for all $a\in[0,1]$.
\end{proof}

\end{document}